\documentclass[twocolumn,A4]{article}
\usepackage[dvips]{graphics}
\begin{document}
\onecolumn
\begin{center}
{\bf{\Large Strange effect of disorder on electron transport through 
a thin film}}\\
~\\
Santanu K. Maiti$^{1,2,*}$ \\
~\\
{\em $^1$Theoretical Condensed Matter Physics Division,
Saha Institute of Nuclear Physics, \\
1/AF, Bidhannagar, Kolkata-700 064, India \\
$^2$Department of Physics, Narasinha Dutt College,
129, Belilious Road, Howrah-711 101, India} \\
~\\
{\bf Abstract}
\end{center}
A novel feature of electron transport is explored through a thin film
of varying impurity density with the distance from its surface. The film, 
attached to two metallic electrodes, is described by simple tight-binding 
model and its coupling to the electrodes is treated through Newns-Anderson 
chemisorption theory. Quite interestingly it is observed that, in the 
strong disorder regime the amplitude of the current passing through the film 
increases with the increase of the disorder strength, while it decreases 
in the weak disorder regime. This anomalous behavior is completely opposite 
to that of conventional disordered systems. Our results also predict that 
the electron transport is significantly influenced by the finite size of 
the thin film.
\vskip 1cm
\begin{flushleft}
\end{flushleft}
\vskip 4.6in
\noindent
{\bf ~$^*$Corresponding Author}: Santanu K. Maiti

Electronic mail: santanu.maiti@saha.ac.in
\newpage
\twocolumn

\section{Introduction}

In the last few decades considerable attention has been paid to the 
propagation of electrons through quantum devices with various geometric
structures where the electron transport is predominantly 
coherent~\cite{nitzan1,nitzan2}. Recent progress in creating such quantum 
devices has enabled us to study the electron transport through them in a 
very tunable environment. By using single molecule or cluster 
of molecules it can be made possible to construct the efficient quantum 
devices that provide a signature in the design of future nano-electronic 
circuits. Based on the pioneering work of Aviram and Ratner~\cite{aviram} 
in which a molecular electronic device has been predicted for the first 
time, the development of a theoretical description of molecular electronic 
devices has been pursued. Later, several experiments~\cite{tali,metz,fish,
reed1,reed2} have been performed through different molecular bridge
systems to understand the basic mechanisms underlying such transport.
Though electron transport properties through several bridge systems
have been described elaborately in lot of theoretical as well as 
experimental papers, but yet the complete knowledge of the conduction 
mechanism in this scale is not well understood even today. For example, 
it is not so transparent how the molecular transport is affected by its 
coupling with the side attached electrodes or by the geometry of the 
molecule itself. Several significant factors are there which control the
the electron conduction across a bridge system and all these effects have
to be considered properly to study the electron transport. In a their
work, Ernzerhof {\em et al.}~\cite{ern2} have manifested a general design 
principle through some model calculations, to show how the molecular 
structure plays a key role in determining the electron transport. The 
molecular coupling with the electrodes is also another important factor 
that controls the current in a bridge system. In addition to these, the 
quantum interference of electron waves~\cite{tagami,mag,lau,baer1,baer2,
baer3,gold,ern1,walc1} and the other parameters of the Hamiltonian
that describe the system provide significant effects in the determination 
of the current through the bridge system. Now in these small-scale devices,
dynamical fluctuations play an active role which can be manifested 
through the measurement of ``shot noise'', a direct consequence of the 
quantization of charge. It can be used to obtain information on a system 
which is not available directly through the conductance measurements,
and is generally more sensitive to the effects of electron-electron 
correlations than the average conductance~\cite{blanter,walc2}. 

In this present paper, we will describe quite a different aspect of 
quantum transport than the above mentioned issues. Using the advanced 
nanoscience and technology, it can be made possible to fabricate a 
nano-scale device where the charge carriers are scattered mainly 
from its surface boundaries~\cite{kou,zho1,zho2,ding1,ding2} and not 
from the inner core region. It is completely opposite to that of a 
traditional doped system where the dopant atoms are distributed uniformly 
along the system. For example, in shell-doped nanowires the dopant atoms
are spatially confined within a few atomic layers in the shell region of
a nanowire. In such a shell-doped nanowire, Zhong and Stocks~\cite{zho1} 
have shown that the electron dynamics undergoes a localization to 
quasi-delocalization transition beyond some critical doping. In other very 
recent work~\cite{ding1}, Yang {\em et al.} have also observed the 
localization to quasi-delocalization transition in edge disordered 
graphene nanoribbons upon varying the strength of the edge disorder. 
From the extensive studies of the electron transport in such systems where
the dopant atoms are not distributed uniformly along the system, it
has been suggested that the surface states~\cite{yu}, surface 
scattering~\cite{cui} and the surface reconstructions~\cite{rurali} may 
be responsible to exhibit several diverse transport properties. 
Motivated by such kind of systems, in this article we consider a 
special type of thin film where disorder strength varies smoothly from 
layer to layer with the distance from the film surface. This system 
shows a peculiar behavior of the electron transport where the current 
amplitude increases with the increase of the disorder strength in the 
limit of strong disorder, while the amplitude decreases in the weak 
disorder limit. On the other hand, for the conventional disordered 
system i.e., the system subjected to uniform disorder, the current 
amplitude always decreases with the increase of the disorder strength. 
From our study it is also observed that the electron transport through 
the film is significantly influenced by its size which reveals the 
finite quantum size effects. Here we reproduce an analytic approach 
based on the tight-binding model to investigate the electron transport 
through the film system, and adopt the Newns-Anderson chemisorption 
model~\cite{new,muj1,muj2} for the description of the electrodes and 
for the interaction of the electrodes with the film.

We organize this paper as follows. In Section $2$, we describe the model 
and the methodology for the calculation of the transmission probability 
($T$) and the current ($I$) through a thin film attached to two metallic 
electrodes by the use of Green's function technique. Section $3$ discusses
the significant results, and finally , we summarize our results in Section 
$4$.

\section{Model and the theoretical description}

This section describes the model and the methodology for the calculation of 
the transmission probability ($T$), conductance ($g$) and the current ($I$) 
through a thin film attached to two one-dimensional metallic electrodes 
\begin{figure}[ht]
{\centering \resizebox*{8cm}{4.5cm}{\includegraphics{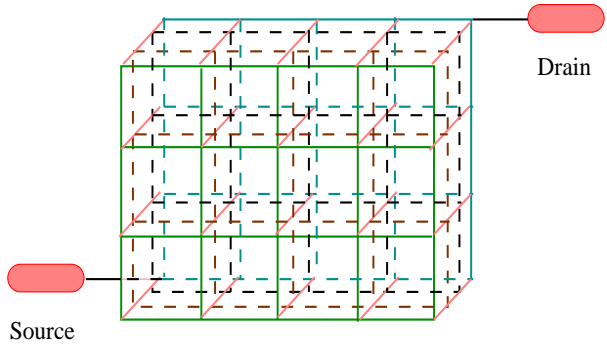}}\par}
\caption{A thin film of four layers attached to two metallic electrodes
(source and drain), where the different layers are subjected to different
impurity strengths. The top most front layer (green color) is subjected to
the highest impurity strength and the strength of the impurity decreases 
smoothly to-wards the bottom layer keeping the lowest bottom layer (light 
blue color) as impurity free. The two electrodes are attached at the two 
extreme corners of the bottom layer.}
\label{quantumfilm}
\end{figure}
by using the Green's function technique. The schematic view of such a 
bridge system is illustrated in Fig.~\ref{quantumfilm}.

For low bias voltage and temperature, the conductance $g$ of the film is 
determined by the Landauer conductance formula~\cite{datta},
\begin{equation}
g=\frac{2e^2}{h} T
\label{equ1}
\end{equation}
where the transmission probability $T$ becomes~\cite{datta},
\begin{equation}
T={\mbox {Tr}} \left[\Gamma_S G_F^r \Gamma_D G_F^a\right]
\label{equ2}
\end{equation}
Here $G_F^r$  and $G_F^a$ correspond to the retarded and advanced Green's 
functions of the film, and $\Gamma_S$ and $\Gamma_D$ describe its coupling 
with the source and the drain, respectively. The Green's function of the 
film is written in this form,
\begin{equation}
G_{F}=\left(E-H_{F}-\Sigma_S-\Sigma_D\right)^{-1}
\label{equ3}
\end{equation}
where $E$ is the energy of the injecting electron and $H_{F}$ represents the 
Hamiltonian of the film which can be written in the tight-binding form within 
the non-interacting picture like,
\begin{equation}
H_{F}=\sum_i \epsilon_i c_i^{\dagger} c_i + \sum_{<ij>} t \left(c_i^{\dagger}
c_j + c_j^{\dagger} c_i\right)
\label{equ4}
\end{equation}
In this expression, $\epsilon_i$'s are the site energies and $t$ corresponds 
to the nearest-neighbor hopping strength. As an approximation, we set the
hopping strengths along the longitudinal and the transverse directions in 
each layer of the film are identical with each other which is denoted by 
the parameter $t$. Similar hopping strength $t$ is also taken between two 
consecutive layers, for simplicity. Now in order to introduce the impurities
in the thin film where the different layers are subjected to different impurity
strengths, we choose the site energies ($\epsilon_i$'s) randomly from a 
``Box" distribution function such that the top most front layer becomes
the highest disordered layer with strength $W$ and the strength gradually
decreases to-wards the bottom layer as a function of $W/N_l$ ($N_l$ be the 
total number of layers in the film), keeping the lowest bottom layer as 
impurity free. On the other hand, in the traditional disordered thin film
all the layers are subjected to the same disorder strength $W$. In our 
present model we use the similar kind of tight-binding Hamiltonian as 
prescribed in Eq.(\ref{equ4}) to describe the side attached electrodes, 
where the site energy and the nearest-neighbor hopping strength are 
represented by the symbols $\epsilon_i^{\prime}$ and $v$, respectively. 
The parameters $\Sigma_S$ and $\Sigma_D$ in Eq.(\ref{equ3}) correspond 
to the self-energies due to coupling of the film with the source and the 
drain, respectively, where all the informations of this coupling are 
included into these two self-energies and are described by the 
Newns-Anderson chemisorption model~\cite{new,muj1,muj2}. This Newns-Anderson 
model permits us to describe the conductance in terms of the effective 
film properties multiplied by the effective state densities involving the 
coupling, and allows us to study directly the conductance as a function of 
the properties of the electronic structure of the film between the electrodes.

The current passing through the film can be regarded as a single electron
scattering process between the two reservoirs of charge carriers. The
current-voltage relationship can be obtained from the expression~\cite{datta},
\begin{equation}
I(V)=\frac{e}{\pi \hbar}\int \limits_{-\infty}^{\infty} \left(f_S-f_D\right) 
T(E) dE
\label{equ5}
\end{equation}
where $f_{S(D)}=f\left(E-\mu_{S(D)}\right)$ gives the Fermi distribution
function with the electrochemical potential $\mu_{S(D)}=E_F\pm eV/2$.
Usually, the electric field inside the thin film, especially for small 
films, seems to have a minimal effect on the $g$-$E$ characteristics. Thus 
it introduces very little error if we assume that, the entire voltage is
dropped across the film-electrode interfaces. The $g$-$E$ characteristics
are not significantly altered. On the other hand, for larger system sizes
and higher bias voltage, the electric field inside the film may play a
more significant role depending on the size and the structure of the
film~\cite{tian}, but yet the effect is quite small.

In this article, we concentrate our study on the determination of the 
typical current amplitude which can be expressed through the relation,
\begin{equation}
I_{typ}=\sqrt{<I^2>_{W,V}}
\label{equ6}
\end{equation}
where $W$ and $V$ correspond to the impurity strength and the applied bias
voltage, respectively.

Throughout this article we study our results at absolute zero temperature,
but the qualitative behavior of all the results are invariant up to some
finite temperature ($\sim 300$ K). The reason for such an assumption is
that the broadening of the energy levels of the thin film due to its 
coupling with the electrodes is much larger than that of the thermal
broadening. For simplicity, we take the unit $c=e=h=1$ in our present
calculations.

\section{Results and discussion}

\begin{figure}[ht]
{\centering \resizebox*{8cm}{11cm}{\includegraphics{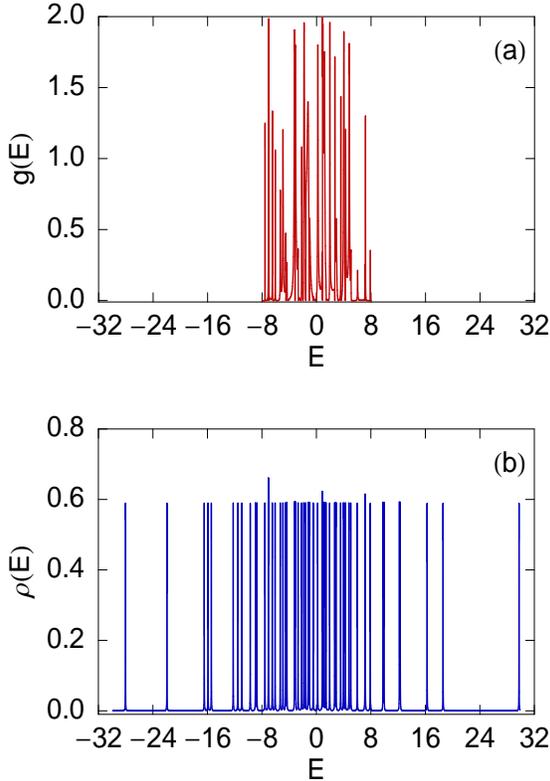}}\par}
\caption{(a) $g(E)$-$E$ (red color) and (b) $\rho(E)$-$E$ (blue color) 
curves for a smoothly varying disordered ($W=30$, strong disorder limit) 
thin film with six layers ($N_z=6$), where the system size of each layer 
is taken as: $N_x=3$ and $N_y=3$.}
\label{conddensity3}
\end{figure}

Here we focus the significant results and describe the strange effect of 
impurity on electron transport through a thin film subjected to the 
smoothly varying impurity density from its surface. These results provide 
the basic conduction mechanisms and the essential principles for the 
control of electron transport in a bridge system. An anomalous feature 
of the electron transport through this system is observed, where the 
current amplitude increases with the increase of the impurity strength 
in the strong impurity regime, while the current amplitude decreases 
with the impurity strength in the weak impurity regime. This peculiar 
behavior is completely opposite to that of the traditional doped film 
in which the current amplitude always decreases with the increase of 
the doping concentration. Throughout our discussion we choose the values 
of the different parameters as follows: the coupling strengths of the 
film to the electrodes $\tau_S=\tau_D=1.5$, the hopping strengths $t=2$ 
and $v=4$ respectively in the film and and in the two electrodes. The 
site energies ($\epsilon_i^{\prime}$'s) in the electrodes are set to 
zero, for the sake of simplicity. In addition to these parameters, 
three other parameters are also introduced those are represented by 
$N_x$, $N_y$ and $N_z$, where the first two of them correspond to the 
total number of lattice sites in each layer of the film along the $x$ 
and $y$ directions, respectively, and the third one ($N_z$) represents 
the total number lattice sites along the $z$ direction of the film. In 
the smoothly varying disordered film, the different layers are subjected 
to the strengths $W_l=W/N_l$, keeping the top most front layer as the 
highest disordered layer with strength $W$ and the lowest bottom layer 
as disorder free. While, for the conventional disordered film, all the 
layers are subjected to the identical strength $W$. Now both for these 
two cases, we choose the site energies randomly from a ``Box"
\begin{figure}[ht]
{\centering \resizebox*{8cm}{5.5cm}{\includegraphics{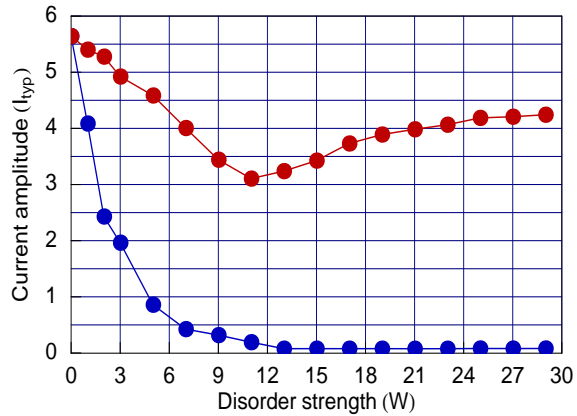}}\par}
\caption{Typical current amplitudes ($I_{typ}$) as a function of the impurity
strength ($W$) for the thin films with six layers ($N_z=6$), where the
system size of each layer is taken as: $N_x=3$ and $N_y=3$. The red and the
blue curves correspond to the results for the smoothly varying and the 
conventional disordered films, respectively.} 
\label{disorder1}
\end{figure}
distribution function, and accordingly, we determine the typical current
amplitude ($I_{typ}$) by averaging over a large number ($50$) of random
disordered configurations in each case to achieve much more accurate result. 
On the other hand, for the averaging over the bias voltage ($V$), we compute
the results considering the range of $V$ within $-16$ to $16$ in each case. 
In this present study, we concentrate ourselves only on the smaller system 
sizes, since all the qualitative behaviors remain invariant even for the 
larger system sizes, and therefore, the numerical results can be computed 
in the low cost of time. 
The variation of the typical current amplitudes ($I_{typ}$) as a function of 
the impurity strength ($W$) for the thin films with system size $N_x=3$, 
$N_y=3$ and $N_z=6$ is shown in Fig.~\ref{disorder1}. The red and the blue 
curves correspond to the results for the smoothly varying and the conventional 
disordered films, respectively. For the conventional disordered film, the 
typical current amplitude decreases sharply with the increase of the impurity 
strength ($W$). This behavior can be well understood from the theory of 
\begin{figure}[ht]
{\centering \resizebox*{8cm}{11cm}{\includegraphics{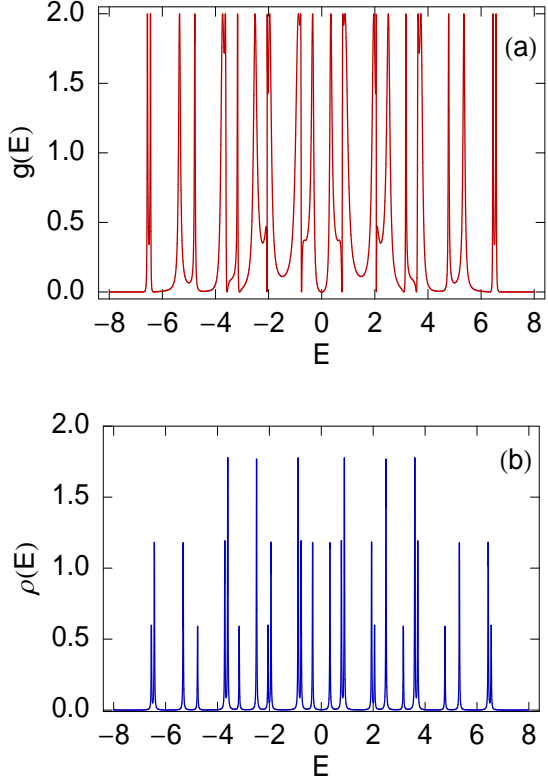}}\par}
\caption{(a) $g(E)$-$E$ (red color) and (b) $\rho(E)$-$E$ (blue color) curves
for an ordered ($W=0$) thin film with six layers ($N_z=6$), where 
the system size of each layer is taken as: $N_x=3$ and $N_y=3$.}
\label{conddensity1}
\end{figure}
Anderson localization, where more localization is achieved with the increase 
of the disorder strength~\cite{lee}. Such a localization phenomenon is 
well established in the transport community from a long back ago. A 
dramatic feature is observed only when the disorder strength decreases 
smoothly from the top most highest disordered layer, keeping the lowest 
bottom layer as disorder free. In this particular system, the current 
amplitude initially decreases with the increase of the impurity strength, 
while beyond some critical value of the impurity strength $W=W_c$ (say) 
the amplitude increases. This phenomenon is completely opposite in nature 
from the traditional disordered system, as discussed above. Such an 
anomalous behavior can be explained in this way. We can treat the smoothly
varying disordered film with ordered bottom layer as an order-disorder
separated film. In such an order-disorder separated film, a gradual 
\begin{figure}[ht]
{\centering \resizebox*{8cm}{11cm}{\includegraphics{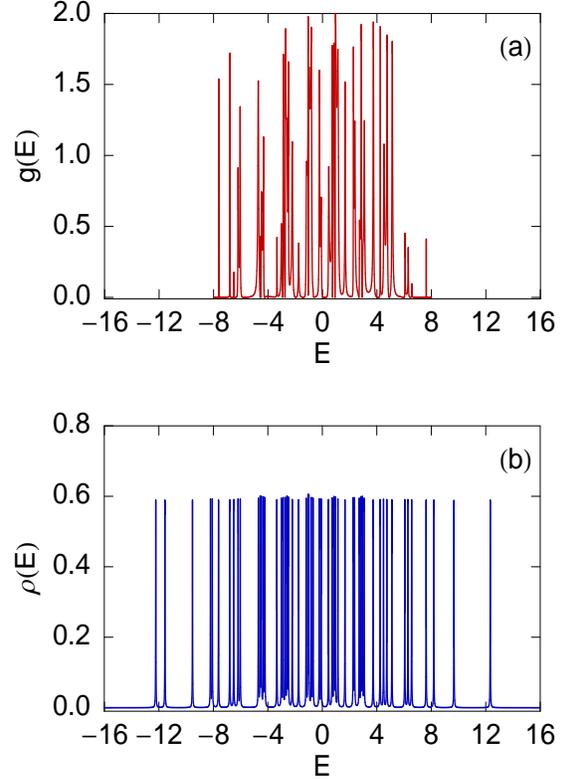}}\par}
\caption{(a) $g(E)$-$E$ (red color) and (b) $\rho(E)$-$E$ (blue color) 
curves for a smoothly varying disordered ($W=10$, weak disorder limit) 
thin film with six layers ($N_z=6$), where the system size of each layer 
is taken as: $N_x=3$ and $N_y=3$.}
\label{conddensity2}
\end{figure}
separation of the energy spectra of the disordered layers and the ordered 
layer takes place with the increase of the disorder strength $W$. In the 
limit of strong disorder, the energy spectrum of the order-disorder
separated film contains localized tail states with much small and central
states with much large values of localization length. Hence the central
states gradually separated from the tail states and delocalized with the
increase of the strength of the disorder. To understand it precisely, here
we present the behavior of the conductance for the three different cases
considering the disorder strengths $W=0$, $W=10$ and $W=30$. The results
are shown in Fig.~\ref{conddensity1}, Fig.~\ref{conddensity2} and
Fig.~\ref{conddensity3}, respectively. In every case the pictures of the
density of states (DOS) are also given to show clearly that with the increase 
of the disorder strength more energy eigenstates appear in the energy 
regimes for which the conductance is zero. Thus the separation of the
localized and the delocalized eigenstates is clearly visible from these
pictures. Hence for the coupled order-disorder separated film, the 
coupling between the localized states with the extended states is strongly 
influenced by the strength of the disorder, and this coupling is inversely 
proportional to the disorder strength $W$ which indicates that the influence
of the random scattering in the ordered layer due to the strong localization
in the disordered layers decreases. Therefore, in the limit of weak 
disorder the coupling effect is strong, while the coupling effect becomes 
less significant in the strong disorder regime. Accordingly, in the limit 
of weak disorder the electron transport is strongly influenced by the 
impurities at the disordered layers such that the electron states are 
scattered more and hence the current amplitude decreases. On the other 
hand, for the strong disorder limit the extended states are less 
influenced by the disordered layers and the coupling effect gradually 
decreases with the increase of the impurity strength which provide the 
larger current amplitude in the strong disorder limit. For large enough 
impurity strength, the extended states are almost unaffected by the 
impurities at the disordered layers and in that case the current is carried 
only by these extended states in the ordered layer which is the trivial 
limit. So the exciting limit is the intermediate limit of $W$. In order 
to investigate the finite size effect on the electron transport,
we also calculate the typical current amplitude for the other two different
system sizes of the thin film those are plotted in Fig.~\ref{disorder2} and 
\begin{figure}[ht]
{\centering \resizebox*{8cm}{5.5cm}{\includegraphics{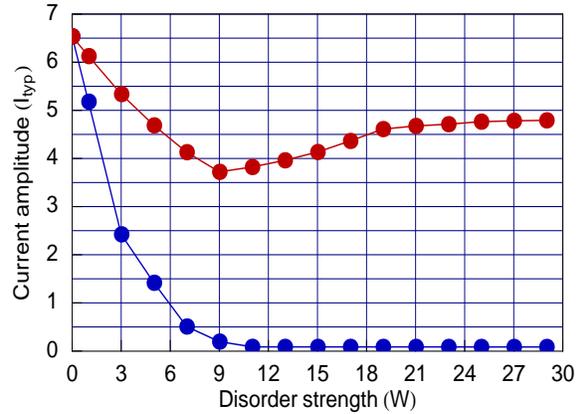}}\par}
\caption{Typical current amplitudes ($I_{typ}$) as a function of the impurity
strength ($W$) for the thin films with seven layers ($N_z=7$), where the
system size of each layer is taken as: $N_x=3$ and $N_y=3$. The red and the
blue curves correspond to the identical meaning as in Fig.~\ref{disorder1}.}
\label{disorder2}
\end{figure}
Fig.~\ref{disorder3}, respectively. In Fig.~\ref{disorder2}, we plot the 
typical current amplitudes for the films with system size $N_X=3$, $N_y=3$ 
and $N_z=7$, while the results for the films with system size $N_x=3$, 
$N_y=3$ and $N_z=8$ are shown in Fig.~\ref{disorder3}. The red and the blue 
curves of these two figures correspond to the identical meaning as in 
Fig.~\ref{disorder1}. Since both for these two films we will get the similar 
behavior of the conductance and the density of states, we do not describe 
these results further. The variation of the typical current amplitudes
for these films with the disorder strength shows quite similar behavior 
as discussed earlier. But the significant point is that the typical current 
amplitude where it goes to a minimum strongly depends on the system size of 
the film which reveals the finite quantum size effects in the study of 
electron transport phenomena. The underlying physics behind the location 
of the minimum in the current versus disorder curve is very interesting. 
The current amplitude is controlled by the two competing mechanisms. One 
is the random scattering in the ordered
\begin{figure}[ht]
{\centering \resizebox*{8cm}{5.5cm}{\includegraphics{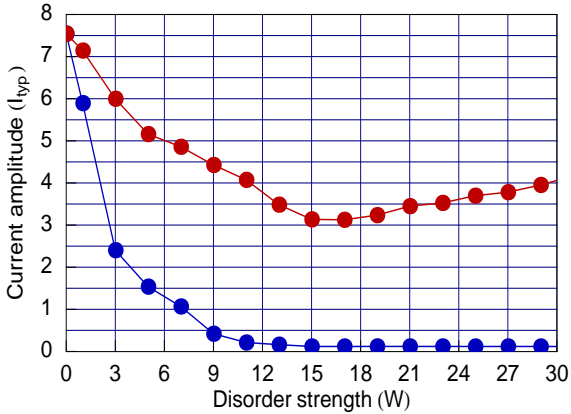}}\par}
\caption{Typical current amplitudes ($I_{typ}$) as a function of the impurity
strength ($W$) for the thin films with eight layers ($N_z=8$), where the
system size of each layer is taken as: $N_x=3$ and $N_y=3$. The red and the
blue curves correspond to the identical meaning as in Fig.~\ref{disorder1}.}
\label{disorder3}
\end{figure}
layer due to the localization in the disordered layers which tends to 
decrease the current, and the other one is the vanishing influence
of random scattering in the ordered layer due to the strong localization
in the disordered layers which provides the enhancement of the current.
Now depending on the ratio of the atomic sites in the disordered region to 
the atomic sites in the ordered region, the vanishing effect of random
scattering from the ordered states dominates over the non-vanishing effect
of random scattering from these states for a particular disorder strength
$(W=W_c)$, which provides the location of the minimum in the current versus
disorder curve.

\section{Concluding Remarks}

In conclusion, we have investigated a novel feature of disorder on electron 
transport through a thin film of variable disorder strength from its surface 
attached to two metallic electrodes by the Green's function formalism. 
A simple tight-binding model has been used to describe the film, where the 
coupling to the electrodes has been treated through the use of Newns-Anderson 
chemisorption theory. Our results have predicted that, in the smoothly 
varying disordered film the typical current amplitude increases 
with the increase of the disorder strength in the strong disorder regime, 
while the amplitude decreases in the weak disorder regime. This behavior is 
completely opposite to that of the conventional disordered film, where the 
current amplitude always decreases with the disorder strength and such a 
strange phenomenon has not been pointed out previously in the literature.
In this context we have also discussed the finite size effect on the electron 
transport by calculating the typical current amplitude in different film 
sizes. From these results it has been observed that, the typical current 
amplitude where it goes to a minimum strongly depends on the size of the 
film which manifests the finite size effect on the electron transport. Thus 
we can predict that, there exists a strong correlation between the localized 
states at the disordered layers and the extended states in the ordered layer 
which depends on the strength of the disorder, and it provides a novel 
phenomenon in the transport community. Similar type of anomalous quantum 
transport can also be observed in lower dimensional systems like, edge 
disordered graphene sheets of single-atom-thick, surface disordered finite 
width rings, nanowires, etc. Our study has suggested that the carrier 
transport in an order-disorder separated mesoscopic device may be
tailored to desired properties through doping for different applications. 

Throughout our discussions we have used several realistic approximations by 
neglecting the effects of the electron-electron interaction, all the inelastic
scattering processes, the Schottky effect, the static Stark effect, etc. More 
studies are expected to take into account all these approximations for our 
further investigations.

\end{document}